\documentclass[a4paper,10pt,aps,prd]{revtex4-2}

\usepackage{lineno}
\usepackage{graphicx,epsf}
\usepackage{amsfonts}
\usepackage{amssymb}
\usepackage{amsmath}
\usepackage{overpic}
\usepackage{subcaption}
\usepackage{fancyhdr}
\modulolinenumbers[5]
\usepackage{framed}
\usepackage{lipsum,color}
\usepackage{amsmath, amsthm, amssymb, amsfonts}
\usepackage{thmtools}
\usepackage{comment}
\usepackage{graphicx}
\usepackage{setspace}
\usepackage{geometry}
\usepackage{float}
\usepackage{hyperref}
\usepackage[utf8]{inputenc}
\usepackage[english]{babel}
\usepackage{framed}
\usepackage[dvipsnames]{xcolor}
\usepackage{tcolorbox}
\usepackage{cancel}

\colorlet{LightGray}{White!90!Periwinkle}
\colorlet{LightOrange}{Orange!15}
\colorlet{LightGreen}{Green!15}

\declaretheoremstyle[name=Theorem,]{thmsty}

\tcolorboxenvironment{theorem}{colback=LightGray}

\declaretheoremstyle[name=Proposition,]{prosty}

\tcolorboxenvironment{proposition}{colback=LightOrange}

\declaretheoremstyle[name=Principle,]{prcpsty}

\tcolorboxenvironment{principle}{colback=LightGreen}

\begin{document}
\title{On the universal content of the proper time flow in scalar and Yang-Mills theories}

\author{Gabriele Giacometti}
\email{gabriele.giacometti@phd.unict.it}
\affiliation{ 
Dipartimento di Fisica e Astronomia “Ettore Majorana”, Università di Catania, Via S. Sofia 64, 95123, Catania, Italy ; \\ 
INFN, Sezione di Catania, Via Santa Sofia 64, 95123 Catania, Italy 
}

\author{Daniele Rizzo}
\email{daniele.rizzo@ncbj.gov.pl}
\affiliation{National Centre for Nuclear Research,
Pasteura 7, 02-093 Warsaw, Poland}

\author{Dario Zappal\`a}
\email{dario.zappala@ct.infn.it}
\affiliation{INFN, Sezione di Catania, Via Santa Sofia 
64, I-95123 Catania, Italy;\\
Centro Siciliano di Fisica Nucleare e 
Struttura della Materia, Catania, Italy.
\vskip 20 pt
}

\date{\today}

\begin{abstract}
\vskip 30pt
\centerline{ABSTRACT}
\vskip 10pt

We investigate the perturbative structure of the proper time
renormalization group flow in scalar and Yang-Mills theories.
Although the PT flow does not belong to the class of exact functional renormalization group equations, 
we show that it correctly reproduces the universal  coefficients of the 
$\beta$-functions at one and two loops.
For the ${\rm O(N)}$ scalar theory, we derive the one- and two-loop 
contributions to the running quartic coupling and also 
confirm the expected anomalous dimension. 
For the ${\rm SU(N)}$ Yang-Mills theory, 
using the background field method, we compute the gauge 
coupling renormalization recovering the correct 
two-loop $\beta$-function 
without generating any gauge-symmetry–violating term. 
These results highlight that, despite its limitations for reconstructing the 
full effective action, the PT flow retains the essential universal content of 
renormalization, accounting for its reliability in diverse applications ranging 
from statistical models to quantum gravity.
\end{abstract}


\maketitle

\section{introduction\label{intro}}

The functional renormalization group (FRG) approach in statistical mechanics 
and quantum field theory inspired by the pioneering work of K. Wilson~\cite{Wilson:1973jj} has achieved a widespread success in the last three 
decades (see e.g.~\cite{Berges:2000ew,Dupuis:2020fhh}), 
providing a powerful tool to study the general renormalization 
properties  of various field theories and to analyze in detail their 
evolution  when the observational (renormalization) scale moves from high 
(ultraviolet (UV)) to low (infrared (IR)) energy ranges. \\
The content of the FRG is enclosed in a 
functional differential flow equation that
describes the evolution of the average effective action 
between the classical action,
taken as the UV (in the limit of the renormalization scale 
$k\to \infty$) boundary condition of the equation, 
and the full effective action (the Legendre transform of the 
generator of the connected Green's functions), which, in principle,
is the solution of the flow equation in the extreme IR region, i.e. 
when $k\to 0$. The details of 
the intermediate steps (finite $k$) of the flow equation are model dependent,
in the sense that they depend on the particular choice of the $k$-dependent 
regulator function that selects the set of modes that are integrated out 
along the flow, respecting
the constraint of recovering, in all cases, the full effective action 
for $k\to 0$~\cite{Wetterich:1992yh,Morris:1993qb,Reuter:1993kw}. Because 
of the latter property, the generation of the perturbative diagrammatic loop 
expansion is naturally produced by the FRG~\cite{Litim:2001ky,Litim:2002xm},
although, on the practical side, replicating perturbative results beyond one 
loop could require a non-trivial effort, as in the simple  
case of the computation of the  two-loop $\beta$-function of
the scalar field~\cite{Papenbrock:1994kf}. \\
The problem of recovering the perturbative scheme from 
a different approach to the flow equation, derived by the original idea 
of generating a flow equation for a  Wilsonian action by means of a 
progressive integration of modes~\cite{Wilson:1973jj,Wegner:1972ih}, 
was also addressed  
~\cite{Liao:1992fm,Liao:1994ip,Liao:1994fp,Bonanno:1996ir,
Bonanno:1997dj,Bonanno:1999ik}, 
and, although the first perturbative order (one-loop) is totally 
under control, it is evident that the nature of the sharp regulator 
acting on momenta, characteristic of this method,
brings in serious flaws when it comes to two-loop computations. \\
The difficulties associated with a sharp momentum cut-off can be somehow 
circumvented by using the Proper Time (PT) formulation of 
the renormalization group flow where the blocking procedure is performed on 
a suitably introduced variable, the proper 
time~\cite{Oleszczuk:1994st,Liao:1994fp,Liao:1995nm,Bonanno:2000yp,Litim:2001hk}. 
The corresponding flow  equation turns out to be
particularly accurate in various contexts, such as the determination of 
critical indices in second order phase 
transition~\cite{Bonanno:2000yp,Mazza:2001bp,Litim:2010tt}, or 
the determination of 
ground state and energy gap of a quanto-mechanical double well 
potential~\cite{Bonanno:2022edf}. It has been largely employed 
in the study of the  gravitational theory~\cite{Bonanno:2004sy, 
Bonanno:2012dg,
Bonanno:2023fij,Bonanno:2025tfj,Bonanno:2025qsc,Bonanno:2025dry}, and 
it has  also been recently related to the essential renormalization 
group~\cite{Falls:2024noj}. \\
Despite this versatility, 
it was shown that the PT flow does not belong to the class of exact 
flows for the effective average action~\cite{Litim:2001hk,Litim:2001ky,Litim:2002xm},
because it is not capable of reproducing the correct loop expansion of the
effective action; rather, it can be regarded as a special case of 
background field flow~\cite{Litim:2002xm,Litim:2002hj}. Along this line, 
the missing elements of the PT flow 
to be assimilated to an exact flow are computed in~\cite{Wetterich:2024ivi}. \\
In this paper, we reconsider this old issue and 
compute the coefficient of the first two orders
of the perturbative series of the $\beta$-function of the 
${\rm O(N)}$ scalar theory and of the ${\rm SU(N)}$ Yang-Mills 
theory from the PT flow, being motivated not only 
by its reliability in many applications, but also by the study in~\cite{deAlwis:2017ysy,Bonanno:2019ukb}, where the PT 
renormalization group equation is regarded  as a 
Wilsonian flow, where, 
at each step of the integration, a new local action is generated,
in contrast with the exact flow that deals with the evolution of 
the full generator of 1-particle irreducible diagrams. \\
In this sense, we can still expect to get significant results at least 
for the determination of  the one- and two- loop  $\beta$-function.
In fact, this is a reduced goal with respect to the reconstruction of the 
full diagrammatic structure of the two-loop effective action
which is only achieved within the class of exact FRG flows, 
but it still gives access to some universal quantities that are
characteristic of each specific theory.
In addition, the computation of the Yang-Mills $\beta$-function allows us 
to verify that the PT  does not generate 
any gauge-violating term (at least in this particular case), 
thus respecting the gauge symmetry of the theory.
Then, in the  next Section the basic details of the calculation
are summarized, while Sec.~\ref{sec3} 
and Sec.~\ref{sec4} are devoted respectively to the $\text{ O(N)}$ theory
and the ${\rm SU(N)}$ Yang-Mills theory. The conclusions are reported in Sec.~\ref{sec5}.

\section{Perturbative expansion and the PT flow}
We start from the regularized  Proper Time (PT) representation of  the 
inverse and the logarithm of an operator $X$ :
\begin{equation}
\label{reginv}
    {X}^{-1}=\int_{1/ \Lambda^2}^{1/k^2}\,ds\,e^{-sX} 
\end{equation}
and
\begin{equation} 
\label{reglog}
  \log[X]=- \int_{1/ \Lambda^2}^{1/k^2} \frac{ds}{s}\; e^{-sX}   \;,
 \end{equation}   
where the integral is performed on the PT variable $s$ and the infinite upper 
limit and zero lower limit of the integrals, 
which define the unregularized original expressions, here are respectively 
replaced with $1/k^2$ and $1/\Lambda^2$, 
and $k$  and $\Lambda$ are two scales  that protect computations from 
possible divergences (IR in the limit $k\to 0$ and UV in the limit $\Lambda \to 
\infty$) 
related to the spectrum of $X$. 
In particular, Eq.~\eqref{reglog} allows us to 
regularize the one-loop effective action  and also obtain a Renormalization
Group (RG) equation for 
the Wilsonian effective action $S_k$, indicated as PT RG flow equation:
\begin{equation}
\label{ptflow}
    k\partial_k \,S_k =-\frac{k \partial_k}{2}  \int_{1/ \Lambda^2}^{1/k^2}  
    \frac{ds}{s}\; {\rm TR}\; \left [ e^{-s\, S_k''} \right] \;,
\end{equation}
where $S_k''$ is the second functional derivative of the 
Wilsonian action $S_k[\phi]$  with respect to its degrees of freedom 
(the field $\phi$), and the trace ${\rm TR}$ is performed 
on all free indices  of  $S_k''$. In particular, in addition to all internal 
indices  associated with $\phi$,  the trace includes the 
integration over the space and momentum variables appearing in $S_k''$.

In order to recover the perturbative series in powers of the coupling,
one can simply replace the running operator $S_k''$ in the right-hand side of Eq.~\eqref{ptflow}, with its value at a fixed order of 
the series, thus obtaining the contribution to $S_k$ to the next order.
To one-loop, this procedure is straightforward. In fact, it is sufficient 
to replace $S_k''$ with the second functional derivative of the 
bare  action which, up to UV counterterms (that are not 
relevant for our purposes), coincides with the tree action $S_0$.
With this replacement, the integration of Eq.~\eqref{ptflow} in the limit  
$1/k^2 \rightarrow \infty $ and $1/\Lambda^2 \rightarrow 0$ reproduces the well 
known one-loop correction 
\begin{equation}
\label{1lcorr}
  S_1 =-\frac{1}{2} \int_0^{\infty}  
    \frac{ds}{s}\; {\rm TR}\; \left [ e^{-s\, S_0''} \right] =   \frac{1}{2} 
    \; {\rm TR}\;  \log \,[S_0''] 
\end{equation}
and the full one-loop action is $S_{1L} =S_0+ S_1 $.
The one-loop $\beta$-function is easily derived by 
following the 
same procedure. In fact, after replacing $S_k''$ with 
$S_0''$ in 
Eq.~\eqref{ptflow}, it is sufficient to pick the coupling 
of interest 
on both sides of the equation with a suitable projector, 
to obtain the 
one-loop flow of the selected coupling.

\begin{figure}
\centering
\includegraphics[scale=.25]{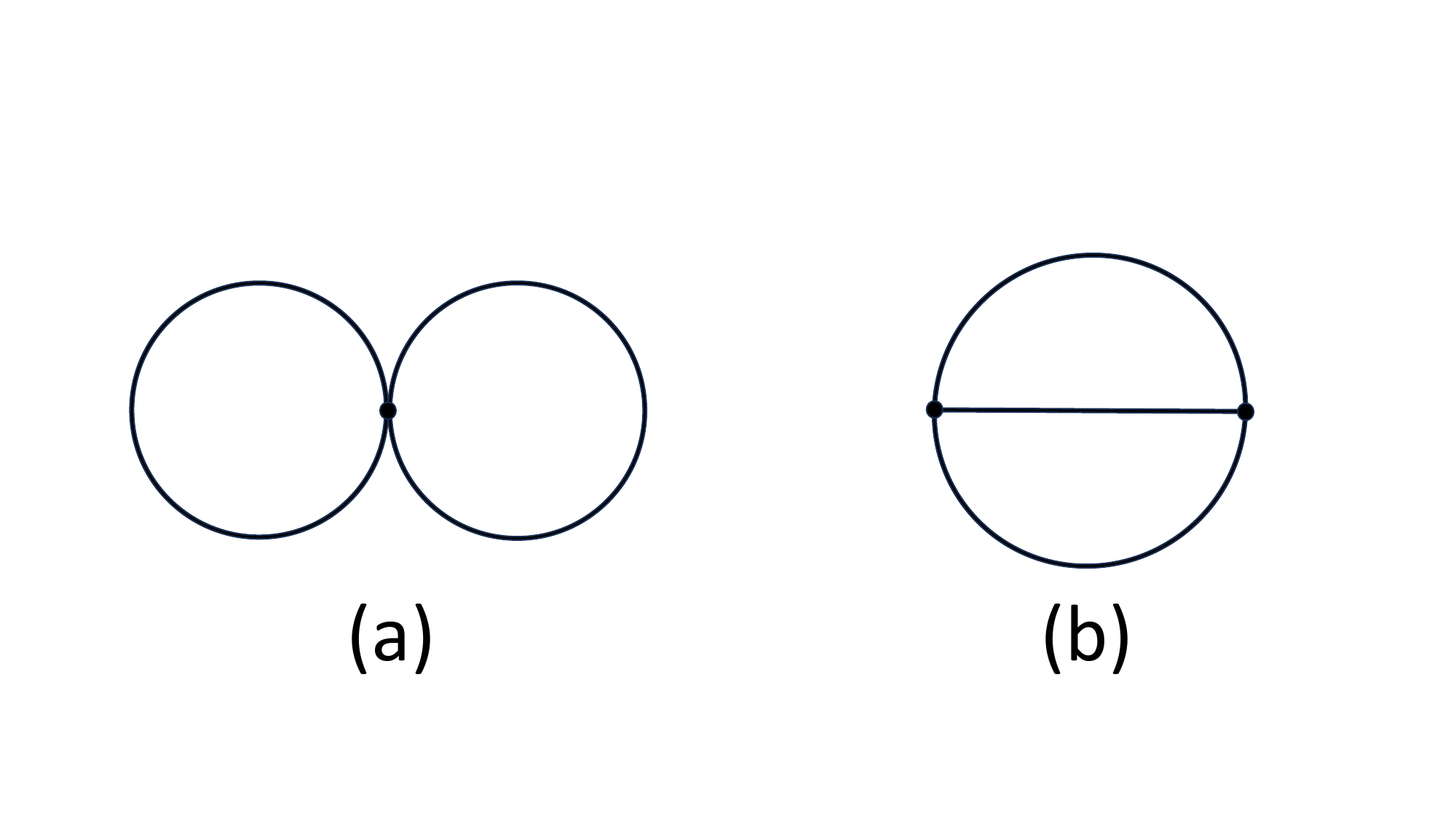}
\caption{Two-loop diagrams contributing to the effective action. Diagram (a) 
is order $\lambda$ and diagram (b) is order $\lambda^2$.}
\label{fig1}
\end{figure}

Unfortunately, the computation of the  
two-loop correction to the effective action, $S_2$, from the straightforward 
extension of Eq.~\eqref{1lcorr},
\begin{equation}
\label{2lcorr}
 S_1+S_2 =-\frac{1}{2} \int_{1/ \Lambda^2}^{1/k^2}  
\frac{ds}{s}\; {\rm TR}\; \left [ e^{-s\, ( S_0''+ S_1'')} 
\right] \;,
 \end{equation}
presents some inconvenience. 
This is due to the structure of the PT flow equation that 
does not possess the structure of the exact flow equation~\cite{Litim:2001ky,Litim:2002xm,Wetterich:2024ivi},
which implies the impossibility of reconstructing the 
complete perturbative structure of the effective 
action. So, for instance, the direct computation of the 
two-loop effective action from the PT 
flow yields a wrong weight for the $O(\lambda^2)$
two-loop vacuum diagram (b) in Fig. \ref{fig1}, as shown in~\cite{Litim:2002xm}.
Nevertheless, we notice that in Eq.~\eqref{2lcorr},
it is still possible to expand at first order the exponential in the 
right-hand side,
\begin{equation}
\label{2lcorrexp}
 S_1+S_2 =-\frac{1}{2} \int_{1/ \Lambda^2}^{1/k^2}  
\frac{ds}{s}\; {\rm TR}\; \left [ e^{-s\,  S_0''}
\left ( 1-s \, S_1''  \right )
\right] \;,
 \end{equation}
so that the zeroth order of the expansion gives 
back the one-loop action $S_1$, 
while the first order produces the two-loop action $S_2$.
But in practice, this formal structure requires the 
full knowledge of the second functional derivative of the 
one-loop action $S_1''$, including its full momentum
dependence. 

Therefore, treating the exponential in 
Eq.~\eqref{1lcorr} as 
a simple function of $S_0''$, 
the second functional derivative of the 
one-loop action
\begin{equation}
\label{1lderiv}
\left( S_1 \right )'' = \left( -\frac{1}{2}  
\int_{1/\Lambda^{2}}^{s}  \frac{dt}{t} \; {\rm TR}\, 
\left [ e^{-t\, S_0''} \right] \right )'' 
= -\frac{1}{2} 
\int_{1/\Lambda^{2}}^{s} \frac{dt}{t} \,  {\rm TR}\,
\left [ e^{-t\, S_0''} \left ( -t \,S_0^{(4)} +t^2 
\, S_0^{(3)} \,S_0^{(3)}\right ) \right] \; ,
\end{equation}
does not actually lead to the complete result for $S_1''$. 
In the latter equation, $S_0^{(3)}$ and $S_0^{(4)}$ indicate respectively three and four functional derivatives of $S_0$ with respect to the field, and the proper time variable is indicated as $t$ to distinguish it from $s$, which instead refers to the proper time variable  of the 'external' loop in Eq.~\eqref{2lcorrexp}.
Nevertheless, it must be remarked that 
Eq.~\eqref{1lderiv} still provides the correct result for 
constant (x-independent) fields, i.e. it can still be retained 
to determine $S_1''$ at zero external momentum.

The computation of the
non-zero external momentum component of $S_1''$
requires instead the use of the following expansion,~\cite{Schwinger:1951nm},
\begin{align}
\label{schw}
&{\rm TR} \left[ 
\,e^{-(A+B)} \right ] = {\rm TR} \Bigg [  \,e^{-A}  - 
B \,e^{-A} + \frac{(-1)^2}{2} \int_0^1 
du_1 \, 
\left (  B\,e^{-(1-u_1) A}  B\,e^{-u_1\,A}   \right ) 
\nonumber\\
&+\;...\;+ \frac{(-1)^{n+1}}{n+1} \int_0^1 du_1 \;...\; 
\int_0^1 du_n \, 
\left(  u_1^{n-1} u_2^{n-2} \;  ...\; u_{n-1}^1 u_n^{0} 
\right )
\nonumber\\ 
&\times B\, \,e^{-(1-u_1) A}  B\, \,e^{-u_1(1-u_2) 
\,A}  B\, \,e^{-u_1 u_2(1-u_3) \,A} \;... \;
B \,e^{- u_1 u_2 \;...\; u_{n-1} u_n \,A} +...\Bigg]
\end{align}
where $A$ depends on some momentum,
while $B$ contains space-dependent fields.
Therefore, the reconstruction of the full momentum dependence of $S_1''$, 
for instance to correctly recover the diagrams in Fig.~\ref{fig1},
does require computing even an infinite sum of terms in the series 
in Eq.~\eqref{schw}. 
This is extremely challenging and beyond the scope of the present paper.

Instead, we recall the indications of~\cite{deAlwis:2017ysy,Bonanno:2019ukb}, 
which suggest that the running action entering the PT flow equation
is a Wilsonian local action. This, in turn, means that the one-loop 
$S_1''$ in the right-hand side of Eq.~\eqref{2lcorrexp} is the second 
derivative of a local object, whose only dependence on the 
square momentum comes from the double functional derivative of its kinetic part.
Therefore, in our computation, we shall treat $S_1$ as a momentum-independent
quantity, except in the presence of an external momentum associated with the two 
functional derivatives of $S_1''$ in  Eq.~\eqref{2lcorrexp}.
This will be the case of the computation of the two-point function at 
two-loop where, with the help of the expansion in Eq.~\eqref{schw}, we shall
retain only the quadratic contributions in the external momentum.

\section{ O(N) scalar theory
\label{sec3}}

We start with the action of the $\rm O(N)$ scalar theory (the index $j$ runs from $1$ to $N$)
\begin{equation}
\label{actionON}
    S_{0}=\int\,d^4x \left[\frac{1}{2}\partial\phi_j\partial\phi_j+\frac{\lambda}{4!}\,(\phi_j \,\phi_j)^2\right]
\end{equation}
and we evaluate the $\beta$-function  of the renormalized coupling $\lambda_R= Z^{-2}\, \lambda$
(the factor $Z$ comes from the field renormalization $\phi_R= Z^{1/2}\phi$) :
\begin{equation}
\label{betadef}
\beta_{\lambda_R} = k\partial_k  \left ( Z^{-2}\, \lambda \right ) = -2  Z^{-3}\, \lambda\,   k\partial_k  (Z)   +  Z^{-2}\,  k\partial_k  \lambda \; ,
\end{equation}
by computing  the $k$-derivative of $Z$ and $\lambda$, order by 
order, directly from the PT flow. \\

By expanding $Z$ in powers of the coupling, $Z=Z_0+Z_1+Z_2+...$, 
with  normalization $Z_0=1$, it is known that $Z_1=0$ if the effective action does not contain vertices with odd powers of the field, 
and therefore,  $Z_1$ does not contribute to  Eq.~\eqref{betadef}. 
As sketched below Eq.~\eqref{1lcorr}, the flow of the coupling $\lambda$
at one-loop order is obtained by replacing 
$S_k''$ with $S_0''$ in   
the right hand-side of
Eq.~\eqref{ptflow}, and then extracting the coefficient of the fourth power of $\phi$ on both sides of the equation,
\begin{equation}
\label{on1l}
\hskip -9pt
\left [ k\partial_k \lambda \right ]_{_{1L}} = -\frac{k 
\partial_k}{2}  
\left .
{\rm P_{_{\phi^4}}}  \left \{ \int_{1/\Lambda^2}^{1/k^2}  
\frac{ds}{s} \int \frac{d^4p}{(2\pi)^4} {\rm Tr} 
\left [ e^{-s \left [ \left ( p^2 + 
\frac{\lambda}{6} \phi_j \phi_j \right ) \delta_{ab} +  
\frac{\lambda}{3} \phi_a\phi_b \right ] } \right]  \right\}  
\right |_{_{\phi=0}} \, ,
\end{equation} 
where we inserted the explicit form of $S_0''$ in the 
exponential and split the full trace of  Eq.~\eqref{ptflow} into 
the product of the integral 
over the four-momentum $p$ times the trace, ${\rm Tr}$, over the 
internal indices associated with the  $\rm N$ field components 
and, finally, 
we introduced the projector ${\rm P_{_{\phi^4}}}=(d^4/(d\phi_i)^4)$ (i.e. the 
fourth derivative with respect to the arbitrary field component $\phi_i$)
to single out the evolution of the coupling $\lambda$.  The projector ${\rm 
P_{_{\phi^4}}}$ can be safely used here, as no non-local effects
are present  in the one-loop calculation  and one can reduce  Eq.~\eqref{on1l} 
to the case of  $x$-independent fields $\phi$, with no loss of generality.

By using  the following  standard result 
$\int \,[ d^4p\; / (2\pi)^4 \;]\, e^{-s\, p^2} = 1/ {(16\pi^2 s^2)}$,
one easily finds 
\begin{equation}
\label{lambda1l}
\left [ k\partial_k \lambda \right ]_{_{1L}} =- 
\frac{\lambda^2}{(16\pi^2)}\;
\frac{(N+8)}{6}
\;k\partial_k\log\frac{\Lambda^2}{k^2} 
\; ,
\end{equation}
and the corresponding  one-loop $\beta$-function:
\begin{equation}
\label{beta1l}
\beta^{1L}_{\lambda_R}= \frac{\lambda^2}{(16\pi^2)}\; \frac{(N+8)}{3} \;.
\end{equation}

Let us now consider the two-loop order. As discussed above, for the 
renormalization of $\lambda$ 
the $x$-dependence of the fields at one-loop 
order can be neglected. Then, Eq.~\eqref{2lcorrexp}  gives
\begin{equation}
\label{on2l}
\left [ k\partial_k \,\lambda \right  ]_{_{2L}}  = 
 - \frac{k \partial_k}{2}  \; {\rm P_{_{\phi^4}}}  
 \left \{ \int_{1/ \Lambda^2}^{1/k^2}  
 \;\frac{ds}{s}\, \int \frac{d^4p_e}{(2\pi)^4}\, {\rm Tr}\; \left [  e^{-s\, S_0''} \, \left (-s \, S_1''  \right ) \right]   \right\}  \Bigg |_{\phi=0} \;.
\end{equation}  
The use of the  projector $ {\rm P_{_{\phi^4}}} $
is allowed because the fields are treated as $x$-independent 
variables. Then, $S_1''$ is computed according to Eq.~\eqref{1lderiv},
\begin{equation}
\label{1lderivapp}
\left( S_1 \right )'' =
 - \frac{1}{2}  \int_{\Lambda^{-2}}^{s} 
\frac{dt}{t}\; \int \frac{d^4p_i}{(2\pi)^4}\, {\rm Tr}\; 
\left [ e^{-t\, S_0''}  \; \left ( -t \,S_0'''' +t^2 \, S_0''' \,S_0'''  \right ) \right] \, ,
\end{equation}
where  $p_i$ is the momentum associated with the 'internal 
loop'  of $S_1$ 
(to be distinguished from $p_e$, which is instead 
associated with the 'external loop' in Eq.~\eqref{on2l}).

After  replacing Eq.~\eqref{1lderivapp} in Eq.~\eqref{on2l}, by applying the projector $ {\rm P_{_{\phi^4}}} $ and integrating on 
the momentum variables
$p_i$ and $p_e$, one finds
\begin{align}
\label{2lpros}
\left [ k\partial_k \,\lambda \right  ]_{_{2L}}  =& \, 
\frac{\lambda^3}{(16\pi^2)^2}
\;  k\partial_k \Bigg\{ \int_{1/\Lambda^{2}}^{1/k^{2}} \frac{ds}{4 s^2}
\int_{1/\Lambda^{2}}^{s} \frac{dt}{t^3} 
\bigg [\frac{t^3 \, (N^2+30N+104) }{9} \\ \nonumber
&+\frac{s\, t^2\, (2 N^2+ 28 N+ 128 ) }{9}+ \frac{ s^2 t\, (N^2+10N+16) }{9}\bigg] \Bigg\} \; .
\end{align}
Since we are interested in the two-loop $\beta$-function, we only look  for terms 
proportional to $\log ( \Lambda/k)$ after the integration in the proper time 
variables $t$ and $s$ in Eq.~\eqref{2lpros}, and therefore we neglect the term 
proportional to $(s\,t^2)$ in square brackets, which produces   $\log^2 ( 
\Lambda/k)$. We neglect as well the  contribution  proportional to $\Lambda^2$, 
coming from the lower extremum of the integration in $dt$ of the last term in 
square brackets, which is to be cancelled by a suitable counterterm to avoid the 
UV divergence. 
The remaining terms in Eq.~\eqref{2lpros} produce 
\begin{equation}
\label{2lend}
\left [ k\partial_k \,\lambda \right  ]_{_{2L}} 
=-\frac{\lambda^3}{(16\pi^2)^2}
\;\frac{(10N+44)}{9} \; .
\end{equation}

Finally, we  evaluate the wave-function renormalization 
$Z$, which enters 
the $\beta$-function computation as indicated in 
Eq.~\eqref{betadef}. $Z$ renormalizes the derivative term
of Eq.~\eqref{actionON}
and  it can be computed from  the derivative of the 
two-point function with 
respect to the  square external momentum $q^2$:
\begin{equation}
\label{zdef}
Z= \left ( \frac{\partial}{\partial\, q^2}  \left [ \frac{\delta^2 S 
}{\delta \widetilde \phi_i (q) \delta \widetilde \phi_i(-
q) } \right ]_{\phi=0} \right )_{q^2=0} \; .
\end{equation}
Note that the index $i$  in Eq.~\eqref{zdef} is not summed, as it simply indicates a 
generic component of the field, and  $Z$ does 
not depend on its particular value. 
Moreover, as mentioned at the beginning of this section, the first non-trivial contribution comes from $S_2$, since the two-point function shows explicit dependence on the external momentum $q$ starting from two-loop order, which is essential to extract a non-zero contribution in Eq.~\eqref{zdef}. 

Then, in order to select the linear dependence in $q^2$ of $S_2''$
we need to use the expansion in Eq.~\eqref{schw} to properly  evaluate the two functional derivatives 
of  $S_1''$,  before  inserting it into Eq.~\eqref{2lcorrexp}.  
Moreover, since  in Eq.~\eqref{zdef}  one has to  set $\phi=0$, 
it is evident that only the term proportional to $B^2$
is actually relevant in the present computation. Therefore, we find
\begin{align}
Z_2=&\Bigg [ \; \frac{\partial}{\partial\, q^2}  
\Bigg (  -\int_{1/ \Lambda^2}^{1/k^2}  \frac{ds}{4} \int 
\frac{d^4p_e}{(2\pi)^4}
\int_{1/ \Lambda^2}^{s} \frac{dt}{t} \int \frac{d^4p_i}
{(2\pi)^4}  \int_0^1\,du \; \frac{\lambda^2 \, (2N+4) }{3}
\nonumber\\
&\times   
e^{-s p_e^2}
\;e^{-t p_i^2(1-u)}
\;e^{-s u (p_i+q)^2}
 \Bigg )\Bigg ]_{q^2=0} \; 
=\; \frac{\lambda^2}{(16\pi^2)^2}
\frac{(N+2)}{36} \; 
\log \frac{\Lambda^2}{k^2}
\label{zfin}
\end{align}
By plugging the results of Eqs.~\eqref{2lend} and~\eqref{zfin} into 
Eq.~\eqref{betadef}, we read the correct 2-loop $\beta$-function for 
the ${\rm O(N)}$ model:
\begin{equation}
\label{beta2l}
\beta_{\lambda_R}^{2L}= \frac{\lambda^3}{(16\pi^2)^2}
 \; \left (   
\frac{(N+2)}{9}   -\frac{(10N+44)}{9}   \right ) = 
- \frac{\lambda^3}{(16\pi^2)^2}
\;  \frac{3N+14}{3} \;.
\end{equation} 

Moreover, since the 2-loop anomalous dimension $\eta$ 
is related to $Z$ by the definition $Z= 1-\eta\; \log 
\left ( {k}/\Lambda \right )$, we find from Eq.~\eqref{zfin}

\begin{equation}
\label{eta}
\eta= \frac{\lambda^2}{(16\pi^2)^2}
\;\;\frac{N+2}{18} \;\;.
\end{equation} 
Eqs.~\eqref{beta1l},~\eqref{beta2l} and~\eqref{eta} 
are the expected  universal renormalization properties of the ${\rm O(N)}$ scalar 
theory at 2-loop level, directly derived from the PT flow equations. \\

Before concluding this Section  we add a couple of comments.  
First, in the above computation  we made use of the PT 
flow in Eq.~\eqref{ptflow}, which can be regarded as 
the limiting case 
of the class of PT flows defined by a 'smooth' regulator 
parametrized by the real number $m>1$ (usually chosen 
as an integer) whose flow, after some manipulations, reads
\begin{equation}
\label{ptsmooth}
    k\partial_k \,S_k =  {\rm TR}\; 
    \left ( 1+ \frac{S_k''}{m\,k^2 } \right)^{-(m+1)} \, ,
\end{equation}
and, in the limit $m\to \infty$, Eq.~\eqref{ptsmooth} 
reproduces the flow in Eq.~\eqref{ptflow}. Then,
it is easy to realize that the one-loop $\beta$-function extracted from
Eq.~\eqref{ptsmooth} coincides  with Eq.~\eqref{beta1l}. However, 
the two-loop $\beta$-function and anomalous dimension from Eq. 
\eqref{ptsmooth} contain additional $O(1/m)$ terms, with
respect to Eqs.~\eqref{beta2l} and~\eqref{eta}, 
that vanish only in  the limit $m\to \infty$.
Therefore, at two-loop order, these universal properties
are only recovered by the PT flow in Eq.~\eqref{ptflow}.

Second, we remark that the two-loop $\beta$-function computation 
of the ${\rm O(N)}$ theory 
by means of the exponential PT flow, was already studied in~\cite{zappalaprd66}
where, instead of working with the full action $S_k$ flow as in the present
case, the PT flow of the two-, four-, and six-point vertices 
is analyzed and the nesting of the one loop diagrams into the vertex 
flow equations is used to evaluate two-loop diagrams. Actually, the two-loop 
$\beta$-function found in~\cite{zappalaprd66}
shows some spurious dependence on ratios of various momenta and of
the running scale $k$, which disappear 
in the limit $k\to 0$ and, consequently, the correct two-loop result 
was obtained by taking  this limit.  Yet, this limit has no physical meaning 
and the correct $\beta$-function should not  show such spurious dependence. 

However,  it is not difficult to show that  in the calculation performed 
in~\cite{zappalaprd66}  the two-loop $\beta$-function
can also  be recovered by keeping $k$ finite and setting instead to 
zero all the 
external momenta of the 
one-loop diagrams, 
when they are inserted in the flow equations to compute the two-loop diagrams
(except for  the anomalous dimension computation where the dependence on the  
external momentum $q$ of the two-point function has to be retained at least to 
order $q^2$, to find a non-vanishing contribution to $Z$ 
according to its definition in Eq.~\eqref{zdef}). 
The latter procedure shares the same spirit of the computation
discussed in this paper, where only the local sector (with constant fields) of 
$S_1$ is used to evaluate $S_2$, again 
expect for the  calculation of the wave 
function renormalization where $O(q^2)$ terms are maintained.
In this sense, our present findings on the two-loop $\beta$-function  fully 
agree with  those of~\cite{zappalaprd66}.

\section{ SU(N) Yang-Mills theory \label{sec4}}
Let us now consider the ${\rm SU(N)}$ Yang-Mills 
theory, defined by the antisymmetric tensor, 
built by means of the gluon field $\widehat 
A_\mu ^a (x)$ ($a=1,2,...,(N^2-1)$) :
\begin{equation}
\label{emtensor}
F_{\mu \nu} ^a (\widehat A) = \partial_\mu 
\widehat A_\nu ^a (x) -  \partial_\nu \widehat 
A_\mu ^a (x) 
+ g f^{abc} \widehat A_\mu ^b (x) \widehat A_\nu 
^c (x) \;,
\end{equation}
where $f^{abc}$ are the totally antisymmetric structure 
constants of the  ${\rm SU(N)}$ group.

Since the use of the background field method~\cite{Abbott} is particularly suitable to 
construct the Wilsonian flow~\cite{deAlwis:2017ysy}, 
we split the  field into  background  and fluctuations
$\widehat A_\mu ^a = A_\mu ^a + Q_\mu ^a$ 
and perform the 
functional integration over the fluctuations $Q_\mu ^a$. 
To this purpose, the background gauge-fixing term is 
introduced 
as the covariant derivative, depending on the 
background field only, applied to the fluctuations
\begin{equation}
\label{bgfixing}
D_\mu^{ac}(A) \; Q_\mu ^c =\Big [\partial_\mu  \delta^{ac}
+ g f^{abc}  A_\mu ^b (x) \Big  ] Q_\mu ^c (x) \;,
\end{equation}
so that the full gauge action, result of the
sum of the Yang-Mills  and gauge-fixing part, is 
\begin{equation}
\label{sa}
{ S}_{A}=
{S}_{\text{YM}} + {S}_{gf}= \int d^4 x' \; \left \{\, 
\frac{1}{4} \, F_{\mu \nu} ^a F_{\mu \nu} ^a +
\frac{1}{2} \, \left ( D_\mu^{ab}(A) \; Q_\mu ^b \right ) 
\; 
\left ( D_\nu^{ac}(A) \; Q_\nu ^c \right ) \, \right \} 
\;\; ,
\end{equation}
where $F_{\mu \nu} ^a$ is a function of the full field 
$\widehat A_\mu ^a (x)$.
The corresponding ghost action is 
\begin{equation}
\label{sg}
S_G=\int d^4 x' \; \left \{\,-\, 
\overline \theta_a
\left (  D^{ab}_\mu(A) \; D^{bc}_\mu(\widehat A)\right ) 
\theta_c 
\, \right \} 
\;\; ,
\end{equation}
and the ghost field is also to be split into background and
fluctuations $\theta_a = \chi_a + c_a$,  in order to 
integrate out the fluctuations  $c_a$.

It is well known that the integration over the 
gluon and ghost fluctuations 
produces, after turning off the ghost background 
$\chi_a=0$, an 
effective action that depends on the background $A_\mu^a$ 
and is invariant under gauge transformations of 
the background~\cite{Abbott}.
However, preserving gauge invariance means that the integration 
of the quantum fluctuations must dress the tree level Yang-Mills 
Lagrangian density by a multiplicative factor, $Z_{_{A}}$,
\begin{equation}
\frac{1 }{4} 
\, F_{\mu \nu} ^a F_{\mu \nu} ^a 
\longrightarrow
\frac{{Z_{_{A}}} }{4} 
\, F_{\mu \nu} ^a F_{\mu \nu} ^a 
\end{equation} 
where $Z_{_{A}}$ is regarded as the background field 
renormalization constant : $A_R =\sqrt{Z_{_{A}}} A$.
Then, since the internal structure of  $F_{\mu \nu}^a (A)$, 
defined in Eq.~\eqref{emtensor} must also be preserved, 
one immediately 
realizes that the quantum corrections to the coupling $g$ 
must be inversely related to those of the background field,
\begin{equation}
\label{grenor}
g_{_R} =  \frac {g}{\sqrt {Z_{_{A}}}} \;,
\end{equation}
where it is understood that $g$ is kept constant, i.e. 
independent of the renormalization scale, while all the 
effects of the renormalization are included in the 
parameter  $Z_{_{A}}$, in contrast with the scalar case where 
both coupling and field are affected by quantum 
corrections in an uncorrelated way.

To determine the loop corrections within the PT flow framework,
it is then sufficient to extract  $Z_{_{A}}$ from the 
analogous versions for the Yang-Mills case 
of  Eqs.~\eqref{1lcorr},~\eqref{2lcorr}. 
The computation of $Z_{_{A}}$ results in a
double expansion in powers of $g^2$ and of 
$\log \left({k^2}/{\Lambda^2}\right)$, as
we discard  terms that vanish in the  limit  $\Lambda\to \infty$:
\begin{equation}
     Z_{_{A}} = 1 + \left [\,\frac{g^2 \,N}{ 16\pi^2 }\;\beta_1 + 
     \left (\frac{g^2 \,N}{ 16\pi^2 } \right )^2    
    \; \beta_2 \,+ O(g^6)\, \right ] \,\log \frac{k^2}
    {\Lambda^2}  + O \left (\log^2 \frac{k^2} 
    {\Lambda^2} \right )\, .
    \label{ZYMexp}
\end{equation}

Then, we get  $\beta_{g_{_R}}$ from  Eq.~\eqref{grenor}  :
\begin{equation}
\label{betaYMdef}
\beta_{g_{_R}} = k\partial_k g_{_R} =
-\;\frac{g}{2\,Z_{_{A}}^{\frac{3}{2}}} \; k\partial_k (Z_{_{A}}) =
- \, g\, \left [\,\frac{g^2 \,N}{ 16\pi^2 }\;\beta_1 + 
     \left (\frac{g^2 \,N}{ 16\pi^2 } \right )^2    
    \; \beta_2 \,+ O(g^6)\, \right ] \; .
\end{equation}
In fact, in computing  Eq.~\eqref{betaYMdef},
we neglect the higher powers of the logarithm
in Eq.~\eqref{ZYMexp}, because they are automatically  
taken into account in $\beta_{g_{_R}}$,  if one  replaces  $g\to g_{_R}$ 
in the right-hand side of Eq.~\eqref{betaYMdef}.
Furthermore, as far as the coefficients $O(g^3)$ and $O(g^5)$ of $\beta_{g_{_R}}$ in  Eq.~\eqref{betaYMdef}
are concerned,  
we can neglect the factor 
$Z_{_{A}}^{-\frac{3}{2}}$ as 
it is proportional to powers of the logarithm.

In computing
the one-loop 
contribution from the PT flow
we cannot use 
Eq.~\eqref{1lcorr}, since in this case the full action 
consists of a bosonic sector, $S_A$,
and a fermionic sector, $S_G$, so that 
the Hessian contains non-diagonal terms.
In fact, by recalling that 
$S_A$ does not contain ghost fields, while $S_G$ is quadratic 
in the ghost field and linear in the gluon 
fluctuations $Q$,  we have: 
\begin{eqnarray}
\label{hessian}
    &&K_{\mu \nu} ^{ab} (A) = \frac{\delta ^2 \;
     S_A (A,Q) }{\delta 
    Q_\mu ^a (x) \;
    \delta Q_\nu ^b (y)} \Bigg|_{Q\,=\,c\,=\,0} 
    \;\;\;\;\;\; ; \;\;\;
    ({{\cal D}_{_{\overline c, Q}}})_{\nu} ^{ab} (A,\chi) = \frac{\delta ^2 \;
    {S}_G (A,Q,\chi,c)}{\delta 
    \overline {c} ^a (x) \;
    \delta Q_\nu ^b (y)} \Bigg|_{Q\,=\,c\,=\,0}
\nonumber\\
    &&({{\cal D}_{_{Q, c}}})_{\mu} ^{ab} (A,\chi) 
    = \frac{\delta ^2 \;
    {S}_G (A,Q,\chi,c)}{
    \delta Q_\mu ^a (x) \;
    \delta {c} ^b (y) } \Bigg|_{Q\,=\,c\,=\,0}
    \;\;\; ; \;\;
        H ^{ab} (A) 
    = \frac{\delta ^2 \;
    {S}_G (A,Q,\chi,c) }    
    {\delta \overline {c} ^a (x) \; c ^b (y)} 
    \Bigg|_{Q\,=\,c\,=\,0}
\end{eqnarray}
with diagonal parts
\begin{eqnarray}
\label{cappa}
K_{\mu \nu} ^{ab} (A) &=& \Big \{  \Big [ -\delta^{ab} \partial_x^2 
+2 g f^{abc}A^c\cdot \partial_x
 +g f^{abc} (\partial_x \cdot A^c)  -g^2 f^{amc} f^{clb} A^m \cdot A^l 
 \Big ]\delta_{\mu\nu}
\nonumber \\
&& +2 g f^{abc} F^c_{\mu\nu}\Big \}\;  \delta^4{(x-y)} 
\end{eqnarray}
and 
\begin{equation}
\label{oghost}
  H ^{ab} (A) = \Big [ -\delta^{ab} \partial_x^2 +2 g f^{abc}A^c\cdot
  \partial_x
+g f^{abc} (\partial_x \cdot A^c)  -g^2 f^{amc} f^{clb} A^m \cdot A^l 
 \Big ]
\;\delta^4{(x-y)} \;.
\end{equation}
Then, the  PT representation requires the computation of  the 
supertrace of the non-diagonal Hessian, which  yields 
the one loop contribution to the Yang-Mills effective action

\begin{equation}
\label{YM1l}
    S_{\text{YM1}} = - \frac{1}{2} \,  \int_{1/ \Lambda^2}^{1/k^2}  
    \frac{ds}{s} \, \text{TR} \left [ e^{-s \, K_{\mu \nu} ^{ab}
    } \right ]
    + \int_{1/ \Lambda^2}^{1/k^2}  
    \frac{ds}{s} \,\text{TR} \left [    e^{-s \, 
    \left[ H^{ab} + M^{ab} \right ]  } \right ]
    \; ,
\end{equation}
where, again, $\text{TR}$ stands for the trace 
on all internal indices and integrals over spacetime and momentum coordinates, 
and the coefficients $-(1/2)$ and $1$ 
in front of the traces respectively provide the weight of 
the bosonic (gluon) and fermionic (ghost) degrees of 
freedom. Finally, the operator $M^{ab}$, generated 
by the non-diagonal part, is
\begin{equation}
\label{misto}
M^{ab} = - 
({{\cal D}_{_{\overline c, Q}}})_{\mu_1} ^{ac} (A,\chi)
\;\cdot [ K^{-1} ]_{\mu_1 \mu_2}^{cd} (A) \; \cdot
({{\cal D}_{_{Q, c}}})_{\mu_2} ^{db} (A,\chi) \; .
\end{equation}
However, at least at one-loop order, we are interested 
in the effective action  $S_{\text{YM1}}$
with zero background ghost field,
$\overline \chi=\chi=0$, and therefore we get no 
contribution from the non-diagonal part as, in this case, 
it is $M^{ab}=0$. The following step is to recover 
$Z_{_{A}}$ from the expansion of 
the exponential in Eq.~\eqref{YM1l} with $ M^{ab}=0$.
We notice that, for the sole purpose of determining $\beta_1$, 
it is sufficient to reconstruct just the $O(A^4)$ part 
of the product $F(A) \, F(A)$, and this can be achieved 
by considering  constant fields $A$, i.e. by neglecting all 
space-derivatives of $A$. Therefore, the expansion of the
exponential is straightforward and 
$Z_{_{A}}$ at order $O(g^2)$ is easily determined. 
Then,  Eqs.~\eqref{ZYMexp},~\eqref{betaYMdef} give
the correct one-loop coefficient of the Yang-Mills
$\beta$-function, as it was already obtained in~\cite{Liao:1995nm},

\begin{equation}
\label{betaYM1l}
 \beta_1= \frac{11}{3} \; .
\end{equation}
Additionally, we verified  that this 
procedure on the one hand does not generate any term 
quadratic  in the fields $O(A^2)$ and,
on the other hand, it does not produce any renormalization 
of the gauge-fixing action $S_{gf}$. This is in agreement 
with the gauge invariance property of $S_{\text{YM1}}$ 
proved in~\cite{Abbott}, 
as the generation of either term
would break gauge symmetry.

So far the computation is quite simple because 
Eq.~\eqref{betaYM1l} only needs the
projection of Eq.~\eqref{YM1l} on constant fields, while
the determination of the full one loop effective action 
$S_{\text{YM1}}$, essential to compute the 
two-loop effects, requires a more careful analysis.
In fact, $S_{\text{YM1}}$ not only contains the field derivative 
contribution to the term proportional to $F^2$
but it also includes novel gauge invariant terms, e.g. 
$O(F^3)$ terms.

Adapting Eq.~\eqref{2lcorrexp} to contain both the bosonic (gluon) and fermionic (ghost) degrees of freedom, it is possible to extract the two-loop effective action of the Yang-Mills theory within PT regularization as 
\begin{equation}
\label{YM2l}
S_{\text{YM2}} = \int_{1/ \Lambda^2}^{1/k^2}  
\frac{ds}{s}\; \left \{ -\frac{1}{2}
{\rm TR}\; \left [ -s \;S_{\text{YM1}}'' \;e^{-s\, K}  \right]
+  {\rm TR}\; \left [ -s \;\ddot{S}_{\text{YM1}} \; e^{-s\, H }
\right]
\right\}
\;,
\end{equation}
where color and Lorentz indices are omitted for simplicity, 
the double prime indicates the double functional 
derivative with respect to the field $A$ while the double dot refers to 
the ghost and antighost functional derivatives. 
We recall that in Eq.~\eqref{YM2l}, the full $S_{\text{YM1}}$ must be
retained in the form given in Eq.~\eqref{YM1l}, without 
setting the ghost background to zero
the subsequent computations. 
Therefore, $S_{\text{YM1}}''$  and $\ddot{S}_{\text{YM1}}$ in Eq.~\eqref{YM2l} must be written 
according to the expansion in Eq.~\eqref{schw}:
\begin{eqnarray}
\label{YM1ldpr}
    S_{\text{YM1}}'' &=& - \frac{1}{2} \,  \int_{1/ \Lambda^2}^{s} 
    \frac{dt}{t} \left\{  \text{TR} 
    \left [ \int_0^1 du \;  t^2
    K' \, e^{-t \,(1-u)  K } \,K' \, e^{-t \,u\,  K } \right ]
    +\;\text{TR} \left [ -t
    K'' \, e^{-t \,K }  \right ]
 \right \} +
 \nonumber \\
     &&   \int_{1/ \Lambda^2}^{s} 
    \frac{dt}{t}  \left\{  \text{TR} 
    \left [ \int_0^1 du \;  t^2
    R' \, e^{-t \,(1-u)  R } \,R' \, e^{-t \,u\,  R } \right ]
    +\;\text{TR} \left [ -t
    R'' \, e^{-t \,R }  \right ] \right \} \;,
\end{eqnarray}

\begin{equation}
\label{YM1lddo}
    \ddot{S}_{\text{YM1}} =   \int_{1/ \Lambda^2}^{s} 
    \frac{dt}{t}
     \left\{  \text{TR} 
    \left [ \int_0^1 du \; t^2
    \dot{R} \, e^{-t \,(1-u)R } \,\dot{R} \, e^{-t \,u R} \right ]
    +\;\text{TR} \left [ -t \;
     \ddot{R} \, e^{-t \,R }  \right ]
 \right \}   
    \; ,
\end{equation}
where we used the compact notation $R=(H+M)$.

Eq.~\eqref{YM2l} is very difficult to 
treat and the reconstruction of the $O(A^4)$ sector of the 
term  $F^2$ is very intricate. Therefore, in order to
minimize the effort in determining $\beta_2$, we exploit the 
relation between the renormalization factor of the field $A$
and of the coupling $g$, valid in the background gauge~\cite{Abbott}.
Then, instead of looking  at the full $S_{\text{YM2}}$, 
we can extract the contribution of the two-loop field renormalization 
from the coefficient of the square external momentum $q^2$ 
in the two-point Green function at two-loop order.
The latter is obtained by taking two functional derivatives of 
$S_{\text{YM2}}$ in Eq.~\eqref{YM2l} and then setting all fields to zero,
analogously to the computation of $Z$ in Eq.~\eqref{zdef}.  
With vanishing fields, the computation becomes much simpler. \\

Moreover, of all the possible ways of performing two functional 
derivatives
in Eq.~\eqref{YM2l}, only the application of both derivatives to 
$S_{\text{YM1}}''$ contributes  to $\beta_2$,
while the derivatives of other field dependent terms, such as 
$\ddot{S}_{\text{YM1}}$  or the exponential factors $e^{-s\, K}$ and 
$e^{-s\, H }$, either give zero contribution to $Z_{_{A}}$ in the limit
$\Lambda \to \infty$, 
or produce terms proportional to $\log^2 (k^2/\Lambda^2)$, 
which are irrelevant to $\beta_2$.
On the other hand, the application of the two functional derivatives 
to $S_{\text{YM1}}''$ in Eq.~\eqref{YM2l} 
produces three relevant contributions.

The first contribution, indicated below  as $\Sigma_1$, 
comes from the application of 
the two derivatives to the non-exponential 
part  of $S_{\text{YM1}}''$ in Eq.~\eqref{YM1ldpr} :
\begin{eqnarray}
\label{sigma1}
    \Sigma_1 =  -  \int_{1/ \Lambda^2}^{s} 
    dt\;t  \int \frac{d^4p_i}{(2\pi)^4}       
    &\Bigg \{ & \text{Tr}  \left [ \int_0^1 du \;   
     \delta [K'] \;  e^{-t \,(1-u) (p_i+q)^2 } \; \delta [K']     
    \; e^{-t \,u\, p_i^2 } \right ]   -
 \nonumber \\ 
     &2& \;\text{Tr} \left [ 
    \int_0^1 du \; \delta [H'] \;  e^{-t \,(1-u) (p_i+q)^2 } \; \delta [H']     
    \; e^{-t \,u\, p_i^2 } \right ]
 \Bigg \}  \;.
\end{eqnarray}
In Eq.~\eqref{sigma1} we indicate
the functional derivatives that define the two-point function with 
a $\delta$, and it is understood that eventually all fields are set 
to zero which, in turn, implies the replacement of $R'$ with $H'$, 
because of the dependence of $M$  on the ghost fields, 
and also implies the suppression of the fields in 
the exponential terms  displayed above. 
The momentum $q$ is the external momentum related to the  
functional derivatives and the trace in the first and second line 
are on different sets of
internal indices, as in the first line it covers both 
color and Lorentz indices, while in the second line 
it refers to color indices only. Finally, in Eq.~\eqref{sigma1}, 
an overall factor 2 is included to count the two equivalent ways 
of performing the functional derivatives.\\

The second contribution, $\Sigma_2$, is obtained by applying both derivatives 
on the exponential part of $S_{\text{YM1}}''$ in Eq.~\eqref{YM1ldpr}:
\begin{eqnarray}
\label{sigma2}
    \Sigma_2  = &-& \int_{1/ \Lambda^2}^{s} 
    \frac{dt}{2t} \int \frac{d^4p_i}{(2\pi)^4} \Bigg \{
      \text{Tr} \left [ \Big( t^4  K' K' - t^3 K'' \Big)      
        \int_0^1 du \, 
    \delta [K] \,  e^{-t \,(1-u) (p_i+q)^2 } \, \delta [K]     
    \, e^{-t \,u\, p_i^2 } \right ]  - 
 \nonumber \\ 
      &&2\,\text{Tr} \left [ \Big( t^4  H' H' - t^3  H'' \Big)      
        \int_0^1 du \,
    \delta [H] \,  e^{-t \,(1-u) (p_i+q)^2 } \, \delta [H]     
    \, e^{-t \,u\, p_i^2 } \right ]  \Bigg \}.
\end{eqnarray} 
In this case, the exponentials in Eq.~\eqref{YM1ldpr} were first 
rearranged into a single factor as they all depend on the same momentum 
(this produces the brackets in front of the integral in the 
variable $u$, both in the first and second line of Eq.~\eqref{sigma2}), 
and then  $\delta [K] $ and  $\delta [H]$ are obtained from the 
double derivation of the exponential factors. Then, as before, 
all fields are set to zero. Note that, in  $\Sigma_2$,
the multiplicity factor 2, due to the two functional derivatives, 
is cancelled by the factor $(1/2)$ coming from the further expansion of 
the exponential, according to Eq.~\eqref{schw}. \\

The third contribution, $\Sigma_3$ is obtained by applying
one derivative on the exponential 
factor, and the other derivative 
on one of the two factors $K'$ (or $R'$) in
Eq.~\eqref{YM1ldpr}. In this case, all possible ways of arranging the two 
derivatives produces a multiplicative factor 4, so that:
\begin{eqnarray}
\label{sigma3}
    \Sigma_3  = &2& \int_{1/ \Lambda^2}^{s} 
    \frac{dt}{t} \int \frac{d^4p_i}{(2\pi)^4} \Bigg \{
      \text{Tr} \left [  t^3 K'     
        \int_0^1 du \; 
     \delta [K] \;  e^{-t \,(1-u) (p_i+q)^2 } \; \delta [K']     
    \, e^{-t \,u\, p_i^2 } \right ]  - 
 \nonumber \\ 
      &2&\,\text{Tr} \left [ t^3  H' \int_0^1 du \;
      \delta [H] \;  e^{-t \,(1-u) (p_i+q)^2 } \; \delta [H']     
    \, e^{-t \,u\, p_i^2 } \right ]  \Bigg \}.
\end{eqnarray} 
Similarly as for the first two contributions, Eq.~\eqref{sigma3} must be evaluated for vanishing fields. \\

When $\Sigma_1$, $\Sigma_2$, $\Sigma_3$ are rearranged within 
Eq.~\eqref{YM2l}, the corresponding coefficients of the external 
momentum $q^2$ produce a contribution to $\beta_2$ in Eq.~\eqref{ZYMexp},
which we indicate as $\beta^{(i)}_2$, with $i=1,2,3$.
We 
find $\beta^{(1)}_2=-5/2$, $\beta^{(2)}_2=53/6$ and 
$\beta^{(3)}_2=5$, 
so that we finally obtain 
the well-known two-loop coefficient of the  Yang-Mills $\beta$-function:
\begin{equation}
\label{betaYM2l}
\beta_2=\beta^{(1)}_2+\beta^{(2)}_2+\beta^{(3)}_2=\frac{34}{3} \;.
\end{equation}

\section{Conclusions \label{sec5}}

The reduction of the PT flow to evaluate the perturbative series that 
defines the $\beta$-function proved successful in determining  the first two 
coefficients of the expansion in powers of the coupling
by means of a compact and rather simple technique,
both for the scalar and the Yang-Mills field theory.
This somehow explains the reliability of the PT flow  predictions  
in several applications. In fact, 
since the  one- and two- loop coefficients of the $\beta$-function  are 
universal, in the sense that they do not depend on the details of the 
regularization procedure at least for mass-independent regularization schemes,
the present computation indicates that the PT flow preserves the main 
physical features associated with the renormalization of these theories. 

Moreover, in the case of the Yang-Mills theory,
we also checked that no gauge symmetry-breaking  'mass' term (i.e. 
proportional to $A^2$),  is  generated in the analyzed quantum corrections. 
On the other hand,  for the scalar theory, no mass term is present 
in the  initial action; had we considered a scalar massive theory with mass
$m$, the universal coefficients of the  $\beta$-function 
would have been modified by $O(m/k)$ corrections, which disappear only for 
$k>>m$ and, in this sense, the PT framework shows the same features of a 
mass-dependent renormalization scheme.

Certainly, the successful computation of one- and two- loop  $\beta$-function 
is not sufficient to warrant the reconstruction of the full perturbative 
structure of the effective action as it was stated in~\cite{Litim:2002xm,Litim:2001ky}. In fact, we verified that such a 
reconstruction  requires a careful expansion and rearrangement
of the exponential structure of the PT flow, and it is not evident at all
if this is realizable in each single diagram. 
In particular, for the Yang-Mills case we are not able to verify that 
a manifestly gauge-invariant effective action is generated
at each order of the perturbative  expansion. 
Fortunately, the latter drawback does not affect the 
two-loop $\beta$-function which is strictly related to the computable  
gauge invariant part of the two-point function.

\begin{acknowledgments}
We are grateful to A. Bonanno and G. Oglialoro for several clarifying 
discussions. The work of DR is supported by the National Science Centre (Poland) under the research grant 2017/26/E/ST2/00470.
\end{acknowledgments}

\bibliography{references}

\end{document}